\documentclass[conference]{IEEEtran}
\IEEEoverridecommandlockouts

\usepackage{cite}
\usepackage{amsmath,amssymb,amsfonts}
\usepackage{graphicx}
\usepackage{textcomp}
\usepackage{xcolor}
\usepackage{comment}
\usepackage[utf8]{inputenc}
\usepackage[margin=1in]{geometry}
\usepackage{tabularx}
\usepackage{hyperref}
\usepackage{multirow} 
\usepackage{mathtools}
\usepackage{url}
\usepackage{caption} 
\usepackage{mathrsfs}
\usepackage{enumitem}
\usepackage{subcaption}
\usepackage{listings}
\usepackage{algorithm}
\usepackage{algorithmic}
\usepackage{xmen}
\usepackage{tikz}
\def\BibTeX{{\rm B\kern-.05em{\sc i\kern-.025em b}\kern-.08em T\kern-.1667em\lower.7ex\hbox{E}\kern-.125emX}}
\renewcommand{\baselinestretch}{1.0}

\usepackage{versions}
\excludeversion{paper}
\includeversion{extended}

\newcommand{\Start}{$\mathit{Start}$}
\newcommand{\Snd}{$\mathit{Snd}$}
\newcommand{\Rcv}{$\mathit{Rcv}$}
\newcommand{\ChannelType}{$\mathit{ChannelType}$}
\newcommand{\insec}{$\mathit{insec}$}
\newcommand{\conf}{$\mathit{conf}$}
\newcommand{\auth}{$\mathit{auth}$}
\newcommand{\secure}{\mathit{sec}}
\newcommand{\step}{\mathit{step}}
\newcommand{\RK}{$\mathit{RK}$}
\newcommand{\G}{$G$}
\newcommand{\qrcode}{$\mathit{qrcode}$}
\newcommand{\vdatalocation}{$\mathit{vdata\_location}$}
\newcommand{\vdatatime}{$\mathit{vdata\_time}$}
\newcommand{\bookingqrcode}{$\mathit{bookingqrcode}$}
\newcommand{\reservationqrcode}{$\mathit{reservationqrcode}$}
\newcommand{\accessqrcode}{$\mathit{accessqrcode}$}
\newcommand{\verificationlink}{$\mathit{verificationlink}$}
\newcommand{\finish}{$\mathit{finish}$}
\newcommand{\vlink}{$\mathit{vlink}$}
\newcommand{\prem}{\mathit{prem}}
\newcommand{\conc}{\mathit{conc}}
\newcommand{\kn}{$\mathit{kn}$}
\newcommand{\UI}{\mathit{UI}}
\newcommand{\normal}{$\mathit{normal}$}

\newcommand{\doublelangle}{$\langle$$\langle$}
\newcommand{\doublerangle}{$\rangle$$\rangle$}

\input{xmen.sty}

\begin{document}

\begin{paper}
\title{User Identification Procedures with Human Mutations: Formal Analysis and Pilot Study}
\end{paper}

\begin{extended}
\title{User Identification Procedures with Human Mutations: Formal Analysis and Pilot Study \\ (Extended Version)}
\end{extended}

\author{
\IEEEauthorblockN{
Megha Quamara and
Luca Vigan\`o}
\IEEEauthorblockA{
\textit{Department of Informatics, King’s College London, London, UK} \\
megha.quamara@kcl.ac.uk, luca.vigano@kcl.ac.uk}
}

\maketitle

\begin{abstract}
User identification procedures, essential to the information security of systems, enable system-user interactions by exchanging data through communication links and interfaces to validate and confirm user authenticity. However, human errors can introduce vulnerabilities that may disrupt the intended identification workflow and thus impact system behavior. Therefore, ensuring the integrity of these procedures requires accounting for such erroneous behaviors. We follow a formal, human-centric approach to analyze user identification procedures by modeling them as security ceremonies and apply proven techniques for automatically analyzing such ceremonies. The approach relies on mutation rules to model potential human errors that deviate from expected interactions during the identification process, and is implemented as the X-Men tool, an extension of the Tamarin prover, which automatically generates models with human mutations and implements matching mutations to other ceremony participants for analysis. As a proof-of-concept, we consider a real-life pilot study involving an AI-driven, virtual receptionist kiosk for authenticating visitors. 
\end{abstract}


\section{Introduction}
\label{Sec:Introduction}

The execution of protocols, such as multi-factor authentication~\cite{barron2021click} and secure messaging~\cite{vaziripour2018action}, intrinsically depends on human users, who may be na\"ive or even trained in some cases. This dependence can complicate protocol analysis, as users might unintentionally deviate from the expected interaction process (e.g., by neglecting or repeating protocol steps). In user identification procedures, which are fundamental to these protocols, such deviations can trigger unintended responses or unpredictable system behavior, such as access denial or incorrect authentication, potentially compromising security. This highlights the need for formal analysis of such procedures to rigorously account for (erroneous) human behavior, independent of attacks, and to ensure that interactions achieve the intended security goals or properties of interest~\cite{vigano2022formal}.

We approach this by modeling user identification procedures as security ceremonies and analyzing them in relation to security goals. A \emph{security ceremony} extends the concept of a security protocol by explicitly including human users who interact with computing systems and exchange messages and data through communication channels, user interfaces, or similar means~\cite{ellison2007ceremony}. This enables us to identify vulnerabilities arising from unexpected or incorrect behavior by human users while interacting with other ceremony participants, such as the system, during the identification process. To this end, we adapt and extend the mutation-based approach presented in~\cite{sempreboni2023mutation}. \emph{Mutations} model potential mistakes the human users might make compared to the behavior specified for them in a ceremony. We consider three mutations of~\cite{sempreboni2023mutation}---\emph{skip}, \emph{add}, and \emph{replace}---and define an additional mutation, \emph{disorder}, for the actions performed in the ceremony. The approach allows mutations in the behavior of other ceremony participants as a consequence of (and to align with) the human-induced mutations. These mutations propagate throughout the ceremony. This facilitates analyzing the original ceremony specification and its possible mutations, including how the ceremony has been (or could be) implemented. If matching mutations result in an attack, the ceremony designers could be consulted to validate the mutated specification and use the attack trace to generate concrete test cases for ceremony implementation. This helps determine whether the attack is a false positive or a real attack. 

To automate our analysis, we use an operational tool-chain support, adapting and extending an existing formal, automated method and tool, X-Men~\cite{sempreboni2020x}, built on top of the Tamarin Prover~\cite{meier2013tamarin}, a tool for automatic, unbounded verification of security protocols. X-Men generates mutated ceremony models, which are then input into Tamarin to identify attacks.

As a proof-of-concept, we apply the approach to a two-factor user identification procedure in a real-life case study of an AI-driven, customer reception kiosk in office buildings. It provides functionalities, e.g., offering general information and facilitating physical access to the building. We identify vulnerabilities arising from visitors' mistakes when interacting with the kiosk and analyze them regarding predefined security goals.

To summarize, the contributions of this paper are (1)~the introduction of a new mutation (\emph{disorder}), (2)~the extension of the X-Men tool, and (3) the pilot study.

We proceed as follows. Section~\ref{Sec:CaseStudy} presents the case study. Section~\ref{Sec:Formalization} details the formalization activity to facilitate analysis in Section~\ref{Sec:Analysis}. Section~\ref{Sec:RelatedWork} reviews related work. Section~\ref{Sec:Conclusions} outlines future work.
\begin{extended} The appendix contains more details on the formalization. \end{extended}

\section{Case Study: \emph{``Receptionist Kiosk''}}
\label{Sec:CaseStudy}

To guide our formal security analysis for user identification procedures, we consider the case study of a \emph{Receptionist Kiosk (RK)}, for visitor reception in office buildings, developed within the framework of the Horizon Europe project ``SERMAS''.\footnote{https://sermasproject.eu/} It integrates a platform-agnostic virtual avatar with an AI-driven chatbot powered by a Large Language Model. The avatar interacts with visitors in a conversational mode to assist with tasks like obtaining information or accessing services.

\subsection{Scenario Description} 
\label{Subsec:ScenarioDescription}

RK will be deployed at Poste Italiane's headquarters in Rome, where visitors, referred to as \emph{guests}, will undergo procedures to ensure safe building access when invited to meet a Poste Italiane employee. In this scenario, RK will perform the following macro-activities: welcome guests, grant access only to those with appointments according to the company's access procedures, and guide them to their destination (e.g., the meeting location). RK can adapt its language to the guest, identify uncertain behaviors, and proactively offer them assistance when needed. Additional actors in this scenario are: \emph{Host}, the Poste Italiane employee who invited the guest, \emph{Booking system} for booking management, and \emph{Access control system} for managing access concerning the turnstiles.

Building access is realized through the following steps (cf.~the sequence diagram in Fig.~\ref{Fig:SequenceDiagram}, with the gray-highlighted part for user identification procedure): 

\begin{figure*}[tp]
\begin{center}
\includegraphics[width=10cm,keepaspectratio]{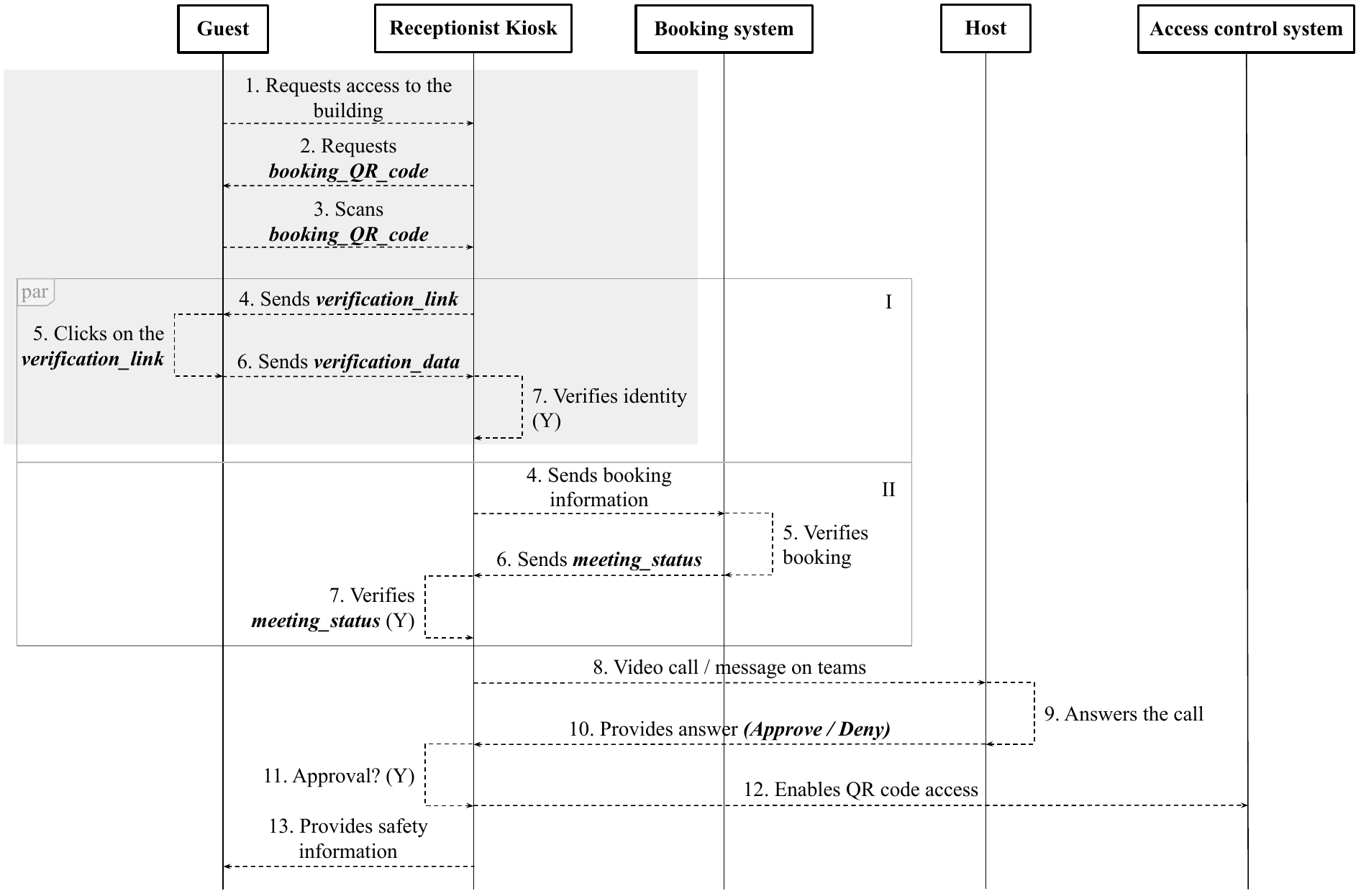}
\end{center}
\caption{Sequence diagram showing RK's expected behavior when a Guest requests building access.} 
\label{Fig:SequenceDiagram}
\end{figure*}

\begin{itemize}
    \item \emph{Reservation:} Guests may visit Poste Italiane's headquarters for various purposes, such as for business meetings or to obtain information about specific services by participating in collaborative sessions. To facilitate these visits and ensure better control, coordination, and security, the host, rather than the guest, initiates the reservation process. He connects to Poste Italiane's web application, enters the required details (guest's name, surname, email, phone number, and visit date/time), and makes the reservation. This sends an invitation email to the guest with a QR code containing the reservation number, required for check-in at the building.
    \item \emph{Arrival at the building (cf. Step 1):} Guest visits Poste Italiane headquarters with their smartphone, which stores the booking QR code and provides access to their email. 
    \item \emph{QR code scanning (cf. Steps 2-3):} At the building entrance, the guest approaches RK, who welcomes them and asks them to scan the booking QR code. Guest uses the kiosk's integrated camera to scan the code from their smartphone.
    \item \emph{Email verification (cf. Steps 4-7(I)) and booking confirmation (cf. Steps 4-7(II)):} After scanning the QR code, RK sends a verification link to the guest's email. The guest confirms their identity by clicking the link, sending a token with meeting details to RK. This serves as a second authentication factor to verify the invited person, while RK simultaneously confirms the meeting status in the booking system.
    \item \emph{Notify the Host (cf. Steps 8-10):} RK notifies the host of the guest's arrival via a video call or a Microsoft Teams message and requests access consent. 
    \item \emph{Access authorization (cf. Steps 11-12):} After successful verification, the booking QR code on the guest's smartphone is activated for building access. The process is repetitive, allowing guest to access the building multiple times a day as long as the access QR code remains valid. 
    \item \emph{Directions for accessing the turnstiles  (cf. Step 13):} RK provides guest with directions to the turnstiles, safety instructions, and guidance to their destination. Additionally, it informs guest of the option to be accompanied by a mobile, robotic kiosk. 
\end{itemize}

\subsection{User Identification Module} 
\label{Subsec:UserIdentificationModule}

The user identification module oversees RK's capability to identify a guest requesting building access based on the provided data (e.g., QR code, verification link, and/or combination of identifiers). Specifically, it addresses the following functional requirements:
\begin{itemize}
    \item \emph{(FR\_1) Guest identification:} RK must identify each guest to ensure secure access to the Poste Italiane building. 
    \item \emph{(FR\_2) Guest authentication:} RK must integrate an authentication procedure via an email link to verify that the person who received the booking QR code is the same individual requesting building access.
\end{itemize}

\section{Formal Specification}
\label{Sec:Formalization}

In this section, we present the formal specification of a security ceremony, an execution model, security goals, and a threat model, with a focus on human mutations. We rely on concepts and notations from existing works on the Tamarin Prover. For brevity, we instantiate only the key concepts and relevant extensions, using the RK's user identification procedure, modeled as UI ceremony.

\subsection{Ceremony Specification}
\label{Subsec:CeremonySpecification}

Formal modeling of a security ceremony involves specifying the behavior of the agents who participate in the ceremony as they execute the ceremony \emph{roles} (that represent specific actions) in order to achieve security goals in an attacker's presence. We consider the following syntactic elements for ceremony specification: 

\begin{itemize}
    \item A \emph{role-script}, which formalizes a role, denotes a sequence of events $e$, characterized by execution of actions $a$ from the set \emph{RoleActions} = \{\Snd, \Rcv, \Start\}. Each event $e$ involves the execution of exactly one action $a$.
    
    \item The \emph{send} and \emph{receive} events respectively take the form \Snd($A$, $l$, $P$, $m$) and \Rcv($A$, $l$, $P$, $m$), where 
    \begin{itemize}
        \item $A$ denotes the name of the role executing the action underlying an event, i.e., send or receive a message,
        
        \item $l$ $\in$ \ChannelType, where $\mathit{ChannelType}$ = \{\insec, \conf, \auth, $\secure$\}, denotes the type of the message transmission channel (i.e., insecure, confidential, authentic, secure), 
        
        \item $P$ denotes the name of the role intended to ``process'' the event, and 
        
        \item $m$ denotes the message being transmitted. 
    \end{itemize}
    
    For the \Snd($A$, $l$, $P$, $m$) event, $P$ denotes the intended recipient of the message $m$. For the \Rcv($A$, $l$, $P$, $m$) event, $P$ denotes the apparent sender of the message $m$ (as the attacker may have forged the identity) and $m$ denotes the expected message (as the attacker may have forged the message).
    
    \item The \emph{start} event takes the form \Start($A$, $K$), where $K$ denotes $A$'s initial knowledge. It initializes a role-script and occurs once during the ceremony.
\end{itemize}

The UI ceremony has 2 roles: the guest $G$ and the receptionist kiosk $RK$ (cf. Fig.~\ref{Fig:SequenceDiagram}). In this ceremony, we do not consider the presence of an explicit attacker. We model this by specifying message transmission over secure channels. Also, we use Tamarin \emph{constants} to define \emph{types} ($t$) in order to identify values exchanged between the roles or those constituting their knowledge base. We represent types within `'. For example, `\qrcode' denotes the type for $\mathit{bookingqrcode}$ or \accessqrcode. This enables restricting the mutations' scope to ensure type safety in that a value can only be replaced by another value of the same type. Thus, the overall transmitted message pattern comprises a constant-value pair $\langle t, m \rangle$.

Fig.~\ref{Fig:RoleScript} shows the role-scripts for the UI ceremony.
\begin{figure}[tbp]
\centering
\resizebox{\columnwidth}{!}{
\begin{tikzpicture}
    \node[text width=11cm, inner sep=0pt] 
    {
    \begin{itemize}
        \renewcommand\labelitemi{}
        \item \textbf{RoleScript$_{G}$} = 
        \\ \text{[}\Start(\G, \doublelangle`\RK', `\qrcode', `\vdatalocation', `\vdatatime'$\rangle$, $\langle$\RK, \bookingqrcode, $\mathit{location}(\mathit{bookingqrcode})$, $time(\mathit{bookingqrcode})$\doublerangle),
        \\ \Snd(\G, $\secure$, \RK, $\langle$`\qrcode', \bookingqrcode$\rangle$), 
        \\ \Rcv(\G, $\secure$, \RK, $\langle$ `\verificationlink', \vlink$\rangle$),
        \\ \Snd(\G, $\secure$, \RK, \doublelangle`\vdatalocation', `\vdatatime', `\verificationlink'$\rangle$, $\langle$$\mathit{location}(\mathit{bookingqrcode})$, $time(\mathit{bookingqrcode})$, \vlink\doublerangle),
        \\ \Rcv(\G, $\secure$, \RK, \doublelangle`\qrcode', `\finish'$\rangle$, $\langle$\accessqrcode, \finish\doublerangle)\text{]} 
        \newline
        \item \textbf{RoleScript$_{\mathit{RK}}$} = 
        \\ \text{[}\Start(\RK, \doublelangle`\G', `\verificationlink'$\rangle$, $\langle$\G, \vlink\doublerangle),
        \\ \Rcv(\RK, $\secure$, \G, $\langle$`\qrcode', \bookingqrcode$\rangle$),
        \\ \Snd(\RK, $\secure$, \G, $\langle$`\verificationlink', \vlink$\rangle$),
        \\ \Rcv(\RK, $\secure$, \G, \doublelangle`\vdatalocation', `\vdatatime, `\verificationlink'$\rangle$, $\langle$$\mathit{location}(\mathit{bookingqrcode})$, $time(\mathit{bookingqrcode})$, \vlink\doublerangle),
        \\ \Snd(\RK, $\secure$, \G, \doublelangle`\qrcode', `\finish'$\rangle$, $\langle$\accessqrcode, \finish\doublerangle)\text{]} 
    \end{itemize}
    };
\end{tikzpicture}}
\caption{Role-scripts for the UI ceremony agents.}
\label{Fig:RoleScript}
\end{figure}

\subsection{Execution Model}
\label{Subsec:ExecutionModel}

We rely on Tamarin's execution model, where protocols are specified using \emph{multiset rewriting rules} to construct a Labeled Transition System (LTS). This model captures evolution in the system's state based on the protocol's actions under execution and the agents' knowledge and behavior. Accordingly:
\begin{itemize}
    \item A system's \emph{state} is a multiset of \emph{facts}: \emph{linear facts} model consumable resources and they can be added to and removed from the system state, and \emph{persistent facts} (prefixed with ``!'') model non-consumable resources, which, once added, can never be removed from the system state.
    \item A system's \emph{initial state} is an empty multiset [ ].
    \item A trace \emph{tr} is a finite sequence of multisets of actions $a$, generated by applying labelled state transition rules of the form
    \begin{equation}
       \prem \xlongrightarrow{a} \conc \tag{Eq. 1} 
    \label{Eq.1}
    \end{equation}
    where $\prem$ denotes the premise and $\conc$ denotes the conclusion. Such a rule can be applied to a state that contains facts matching $\prem$. This removes the matching linear facts from the state, adds the instantiated facts from $\conc$ to the state, and records the instantiated actions from $a$ in the trace. These instantiations move the system to a new state.

    An example trace of actions in the UI ceremony involving a guest $G$ and the receptionist kiosk $\mathit{RK}$, interacting over a secure channel $\secure$, is as follows:
    \begin{center}
        ...\\
        \Snd(\G, $\secure$, \RK, \bookingqrcode)\\
        \Rcv(\RK, $\secure$, \G, \bookingqrcode)\\
        \Snd(\RK, $\secure$, \G,  \vlink)\\
        ...
    \end{center}
\end{itemize}

A protocol model comprises \emph{agent rules} specifying the agents' state transitions and communication. For every event $e$ in the role-script of an agent $A$, the transition rule in~\ref{Eq.1} is structured as follows: 
\begin{itemize}
    \item $\prem$ contains an agent state fact AgSt($A$, $\step$, \kn) and $\conc$ contains the subsequent agent state fact AgSt($A$, $\step$, $\mathit{kn'}$), where $\step$ refers to the role step the agent is in and $kn$ is the agent’s knowledge at that step; 
    \item if event $e$ $\in$ $a$ is \Start($A$, $m$), then it is translated to a setup rule, where $conc$ contains the initial agent state AgSt($A$, $0$, $m$);
    \item if $e$ is \Snd($A$, $l$, $P$, $m$), then $\conc$ additionally contains an outgoing message fact Out$_l$($A$, $P$, $m$);
    \item if $e$ is \Rcv($A$, $l$, $P$, $m$), then $\prem$ contains an incoming message fact In$_l$($P$, $A$, $m$).
\end{itemize}
An agent's knowledge increases monotonically during the ceremony, especially upon receiving a message.

\begin{paper}
The agent rules for the guest role in the UI ceremony are shown in Fig.~\ref{Fig:AgentRulesGuest}, and those for the receptionist kiosk role are provided in~\cite{QuamaraVigano25}.
\end{paper}
\begin{extended}
The agent rules for the guest role in the UI ceremony are shown in Fig.~\ref{Fig:AgentRulesGuest}, and those for the receptionist kiosk role are provided in Appendix~\ref{App:A}.
\end{extended}

\begin{figure}[tbp]
    \centering
    \resizebox{\columnwidth}{!}{
    \begin{tikzpicture}
        \node[text width=11cm, inner sep=0pt] 
        {
            \begin{itemize}[label=$G\_\arabic$]
                \item [(G0)] \text{[ ]}
                \\ $\xlongrightarrow{\substack{\text{\Start(\G, \doublelangle`\RK', `\qrcode', `\vdatalocation', `\vdatatime'$\rangle$,} \\\text{$\langle$\RK, \bookingqrcode, $\mathit{location}(\mathit{bookingqrcode})$, $time$($\mathit{bookingqrcode})$\doublerangle)}}}$
                \\ \text{[}AgSt(\G, 1, $\langle$\RK, \qrcode, \vdatalocation, \vdatatime$\rangle$)\text{]} 
                \newline
                \item [(G1)] \text{[}AgSt(\G, 1, $\langle$\RK, \qrcode, \vdatalocation, \vdatatime$\rangle$)\text{]}
                \\ $\xlongrightarrow{\text{\Snd(\G, $\secure$, \RK, $\langle$`\qrcode', \bookingqrcode$\rangle$)}}$ 
                \\ \text{[}AgSt(\G, 2, $\langle$\RK, \qrcode, \vdatalocation, \vdatatime$\rangle$), 
                Out$_{sec}$(\G, \RK, $\langle$`\qrcode', \bookingqrcode$\rangle$)\text{]}
                \newline
                \item [(G2)] \text{[}AgSt(\G, 2, $\langle$\RK, \qrcode, \vdatalocation, \vdatatime$\rangle$), In$_{sec}$(\RK, \G, $\langle$`\verificationlink', \vlink$\rangle$)\text{]}
                \\ $\xlongrightarrow{\text{\Rcv(\G, $\secure$, \RK, $\langle$`\verificationlink', \vlink$\rangle$)}}$ 
                \\ \text{[}AgSt(\G, 3, $\langle$\RK, \qrcode, \vdatalocation, \vdatatime, \verificationlink$\rangle$)\text{]} 
                \newline
                \item [(G3)] \text{[}AgSt(\G, 3, $\langle$\RK, \qrcode, \vdatalocation, \vdatatime, \verificationlink$\rangle$)\text{]}
                \\ $\xlongrightarrow{\substack{\text{\Snd(\G, $\secure$, \RK, \doublelangle`\vdatalocation', `\vdatatime', `\verificationlink'$\rangle$,} \\ \text{$\langle$$\mathit{location}(\mathit{bookingqrcode})$, $\mathit{time}(\mathit{bookingqrcode})$, \vlink\doublerangle)}}}$
                \\ \text{[}AgSt(\G, 4, $\langle$\RK, \qrcode, \vdatalocation, \vdatatime, \verificationlink$\rangle$), Out$_{sec}$(\G, \RK, \doublelangle`\vdatalocation', `\vdatatime', `\verificationlink'$\rangle$, $\langle$$location(\mathit{bookingqrcode})$, $time(\mathit{bookingqrcode})$, \vlink\doublerangle)\text{]} 
                \newline
                \item [(G4)] \text{[}AgSt(\G, 4, $\langle$\RK, \qrcode, \vdatalocation, \vdatatime, \verificationlink$\rangle$), In$_{sec}$(\RK, \G, \doublelangle`\qrcode', `\finish'$\rangle$, $\langle$\accessqrcode, \finish\doublerangle)\text{]}
                \\ $\xlongrightarrow{\substack{\text{\Rcv(\G, $\secure$, \RK, \doublelangle`\qrcode', `\finish'$\rangle$, $\langle$\accessqrcode, \finish\doublerangle), $Gfin$(\G, `\qrcode',} \\ \text{\accessqrcode)}}}$
                \\ \text{[ ]}
            \end{itemize}
        };
    \end{tikzpicture}}
    \caption{Agent rules for agent $G$ in UI ceremony.}
    \label{Fig:AgentRulesGuest}
\end{figure}

\subsection{Security Goals}
\label{Subsec:Security Goals}

Security goals express the security properties a ceremony should guarantee. In our case, we ``pushed'' security from the system to the human users (i.e., the guests), as human errors during interaction can jeopardize security. We show this with the following three goals (concerning \emph{FR\_1} and \emph{FR\_2}) that we analyze for the UI ceremony:
\begin{itemize}
    \item \emph{(SG\_1)} The guest can obtain the access QR code only via the verification link; i.e., only authenticated guests should obtain the access QR code. 
    
    \item \emph{(SG\_2)} The guest obtains the access QR code corresponding to a valid booking. 
    
    \item \emph{(SG\_3)} The guest cannot use two different booking QR codes to access the building. 
\end{itemize}

\subsection{Threat Model}
\label{Subsec:ThreatModel}

We adapt the threat model proposed in~\cite{sempreboni2019threats} to the UI ceremony. It comprises the following four components:
\begin{enumerate}
    \item \emph{Agents} that take part in a security ceremony. We consider technical systems and humans as the agents, where the latter interact with the former. A technical system works by executing a program. Depending on the level of abstraction of the system representation in a given ceremony, a technical system could be either a piece of hardware/software, a sub-system, or the system as a whole. This extends the architecture-level focus in~\cite{sempreboni2019threats} to also include a high-level system perspective. We envision one or more technical systems and humans participating in the ceremony. Formally, the agents of a ceremony $C$ are specified as follows:
    \begin{center}
         \emph{Technical}($C$) := \{$\mathit{TechSystem_i}$ $\mid$ $i \in \mathbb N$\} \\
         \emph{Human}($C$) := \{$\mathit{Human_j}$ $\mid$ $j \in \mathbb N$\} \\
         \emph{Agent}($C$) :=  $\emph{Technical}$($C$) $\cup$ $\emph{Human}$($C$)
    \end{center}

    The UI ceremony has the following agents:
    \begin{center}
        \emph{Technical}($\UI$) := \{\RK\}\\
        \emph{Human}($\UI$) := \{\G\}\\
        \emph{Agent}($\UI$) := $\emph{Technical}$($\UI$) $\cup$ $\emph{Human}$($\UI$) = \{\G, \RK\}  
    \end{center}

    \item \emph{Information} that the ceremony agents exchange. Information could be of various types, such as tokens or data (e.g., URLs, binary messages), as described in~\cite{sempreboni2019threats}. Additionally, we extend this by considering that information could be a function of the input received through a receive action. Formally, we capture the heterogeneity of the information exchanged among agents in a ceremony $C$ by introducing a free type \emph{Information}($C$). In the UI ceremony,
    \begin{center}
        \emph{Information}($\UI$) := \{\qrcode, \verificationlink, \vdatalocation, \vdatatime\}
    \end{center}
    
    \item \emph{Actions} that each agent performs with other agent(s) by exchanging information. For example, sending messages by technical systems to humans, or humans giving commands to the technical systems. Formally, the set of technical and human actions can be defined as the following ternary relations:
    \begin{center}
        \emph{TechnicalAction}($C$) := $\mathcal{A}_{C,\mathit{TS}}$(\emph{Technical}, \emph{Information}, \emph{Agent})\\
        \emph{HumanAction}($C$) := $\mathcal{A}_{C,H}$(\emph{Human}, \emph{Information}, \emph{Agent})\\
    \end{center}

    The set of actions in a ceremony $C$ is defined as the union of the set of actions of its agents:
    \begin{center}
        \emph{Actions}($C$) := \emph{TechnicalAction}($C$) $\cup$ \emph{HumanAction}($C$)
    \end{center}
    
    \item \emph{Scenarios} that denote the combination of agents and actions, and are characterized by their labeling. Technical and human actions can be labelled as per the Tables~\ref{Tab:TechnicalActionsLabels} and~\ref{Tab:HumanActionsLabels}, respectively. 

    \begin{table}[htbp]
    \centering
    \caption{Labels of technical actions.}
    \label{Tab:TechnicalActionsLabels}
    \resizebox{\columnwidth}{!}{%
    \begin{tabular}{|c|l|}
    \hline
    \textbf{Label} & \multicolumn{1}{c|}{\textbf{Description}}
    \\ \hline
    \emph{normal} & \begin{tabular}[c]{@{}l@{}}Actions prescribed by the ceremony, i.e., analysing a \\ received message and generating another one to send out\end{tabular}
    \\ \hline
    \emph{bug} & \begin{tabular}[c]{@{}l@{}}An unwanted technical deviation from normal, occurring \\ without a specific goal and normally, unexpectedly\end{tabular}
    \\ \hline
    \end{tabular}%
    }
    \end{table}

    \begin{table}[htbp]
    \centering
    \caption{Labels of human actions.}
    \label{Tab:HumanActionsLabels}
    \resizebox{\columnwidth}{!}{%
    \begin{tabular}{|c|l|}
    \hline
    \textbf{Label} & \multicolumn{1}{c|}{\textbf{Description}}
    \\ \hline
    \emph{normal} & Actions prescribed by the ceremony
    \\ \hline
    \emph{error} & \begin{tabular}[c]{@{}l@{}}An unwanted human deviation from normal, occurring \\ without a specific goal and normally, unexpectedly\end{tabular}
    \\ \hline
    \end{tabular}%
    }
    \end{table}

    Thus, the model covers potential attacks through labeled actions, specifically those labeled as bugs or errors, by the technical systems or humans in the ceremony. An agent can be labeled as, for example, \emph{normal}, if it is a technical system (or a human) and all its actions are labeled as normal. We have identified 4 relevant groups of action $a$ labels that an agent $p$ may use, and we define them formally as a set of the following scenarios:
    \begin{center}
        $\mathit{scenario_1}$($p$) := $p \in \emph{Technical}$($C$) and $\forall a \in \emph{TechnicalAction}$($C$, $p$), \emph{label}($a$) = \normal \\
        $\mathit{scenario_2}$($p$) := $p \in \emph{Human}$($C$) and $\forall a \in \emph{HumanAction}$($C$, $p$), \emph{label}($a$) = \normal \\
        $\mathit{scenario_3}$($p$) := $p \in \emph{Technical}$($C$) and $\exists a \in \emph{TechnicalAction}$($C$, $p$), \emph{label}($a$) = $\mathit{bug}$ \\
        $\mathit{scenario_4}$($p$) := $p \in \emph{Human}$($C$) and $\exists a \in \emph{HumanAction}$($C$, $p$), \emph{label}($a$) = $\mathit{error}$ \\
    \end{center}
\end{enumerate}

In the UI ceremony, the most general case in which every agent gets all possible labels observes:
\begin{center}
    $n$($\emph{PLHuman}_{\UI}$(\G)) = 2\\
    $n$($P\emph{LTechnical}_{\UI}$(\RK)) = 2
\end{center}

Hence, the complete threat model chart has width $n$($\emph{Agent}$($\UI$)) = 2 and depth $n$($\emph{PLHuman}_{\UI}$(\G)) * $n$($\emph{PLTechnical}_{\UI}$(\RK)) = 2$^2$ = 4 (cf. Table~\ref{Tab:ThreatModelUICeremony}). In the UI ceremony, we focus on modeling human mistakes with normal system functioning, captured by \#3 in the threat model chart and discussed in the following sub-section.

\begin{table}[htbp]
\centering
\caption{Threat model chart for the UI ceremony with one technical system and one human agent.}
\label{Tab:ThreatModelUICeremony}
\small{%
\begin{tabular}{|c|c|c|}
\hline
\textbf{\#} & \emph{TechSystem} & \emph{Human} 
\\ \hline
1 & \emph{normal} & \emph{normal} 
\\ \hline
2 & \emph{bug} & \emph{normal} 
\\ \hline
3 & \emph{normal} & \emph{error} 
\\ \hline
4 & \emph{bug} & \emph{error} 
\\ \hline
\end{tabular}%
}
\end{table}

\subsection{Modeling Human Mutations}
\label{Subsec:HumanMutations}

Human mutations during a ceremony execution can affect the behavior of other agents and, thus, the whole ceremony. Hence, we define both (1) the human mutations that modify a ceremony trace by mutating the subtrace of the human agent, and (2) the mutations on the subtrace(s) of the other agent(s) that are required to match the human mutations. This results in a mutated ceremony trace, which we feed into Tamarin for execution and analysis concerning the security goals.
Formally, 
\begin{itemize}
    \item a generic \emph{human mutation} is a function
        \begin{equation}
            \mu^H : tr \rightarrow tr' \tag{Eq. 2}
        \label{Eq.2}
        \end{equation}
        that takes as input a trace $tr$ and gives as output a new trace $tr'$ = $[\![tr]\!]^{\mu^H}$ obtained by mutating $H$’s subtrace due to the human $H$ ``deviating'' from the original role-script,
    \item a \emph{matching mutation} $\mu^m$ for a human mutation modifies the sub-traces of other agents to ``match'' and propagate the human mutation $\mu^H$, and
    \item the combination $\mu^H \circ \mu^m$ : $tr \rightarrow tr'$ of the two mutations takes as input a trace $tr$ and gives as output a new trace $tr'$ = $[\![tr]\!]^{\mu^{H} \circ \mu^{m}}$ in which the human mutation is matched and propagated.
\end{itemize}

We denote the sub-trace of a human agent $H$ in a ceremony execution as [$a_0$,$\ldots$, $a_i$,$\ldots$, $a_n$]$^H$, where 0 $< i \leq n$, and $[\![]\!]^\mu$ apply not just to traces but also indirectly apply to the role-scripts and Tamarin actions.

We consider the following four mutations: 
\begin{enumerate}
    \item \emph{Skip}: Involves skipping of action(s) the ceremony expects the human user to carry out.
    \item \emph{Add}: Involves addition of action(s), where the user can (1) send a message at any time or (2) duplicate a \emph{send} action, executing the action twice.
    \item \emph{Replace}: Involves a human user replacing a message being sent by another. This is carried out by combining \emph{skip} and \emph{add} mutations.
    \item \emph{Disorder}: Involves a human user confusing the order of the ceremony actions, e.g., by swapping an action with another. This is also carried out by combining the \emph{skip} and the \emph{add} mutations. However, it differs from replace mutations, which alter message content. Considering disorder mutation in user identification procedures is essential as it captures realistic user behavior, particularly for non-blocking ceremony actions, such as verification or confirmation. For example, in the UI ceremony, a guest might attempt to use an old booking QR code from a previous visit or access a turnstile without scanning a new access QR code. These actions, which do not enforce strict ordering, allow swapping steps without halting the ceremony’s progress.

    Algorithm~\ref{Alg:Disorder} shows the disorder mutation with reference to sequence diagram~\ref{Fig:DisorderMutation}.

    \begin{figure}[tbp] 
    \centering
    \includegraphics[width=6cm,keepaspectratio]{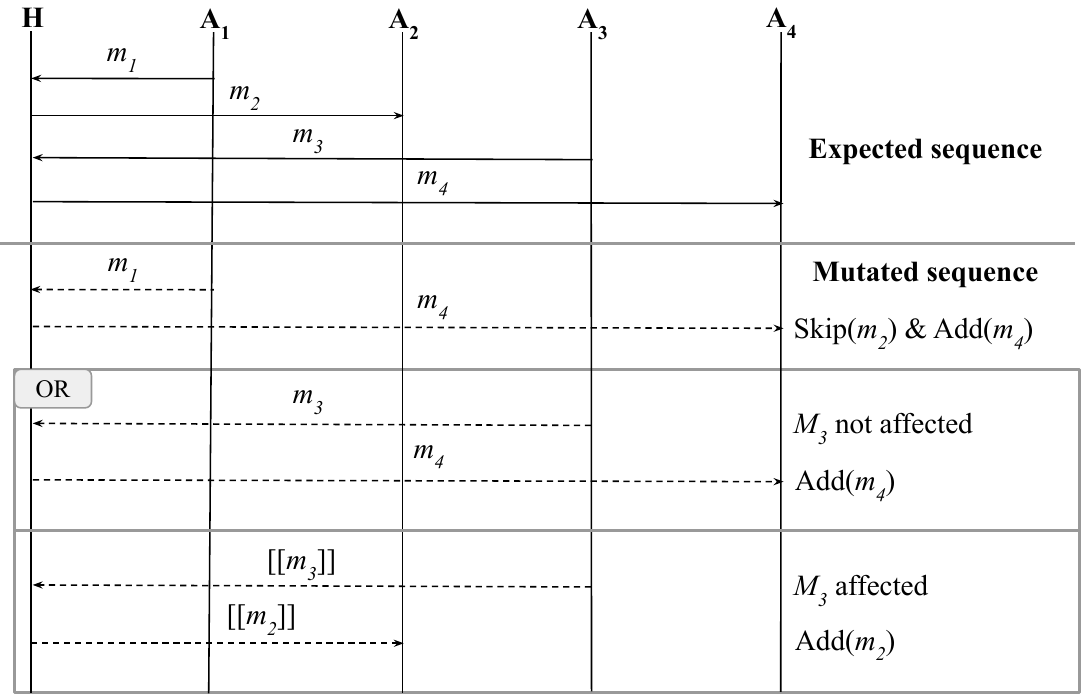}
    \caption{Sequence diagram showing the impact of the \emph{disorder} mutation.} 
    \label{Fig:DisorderMutation}
    \end{figure}

    \begin{algorithm}
\caption{Disorder mutation $\mu_{disorder}^H$}
\label{Alg:Disorder}
\begin{algorithmic}[1]
\STATE build all transitions $i$ obtained by swapping $m_2$ with $m_4$ that is in the powerset of $H$’s current knowledge $kn_{i+1}$, preserving types as specified by the corresponding constants, i.e.,
\STATE AgSt($H$, $i$, \kn), Pre$_i$, \Rcv($H$, $l_1$, $A_1$, $m_1$) $\longrightarrow$ AgSt($H$, $i+1$, $\mathit{kn_{i+1}}$), Post$_{i+1}$, \Snd($H$, $l_4$, $A_4$, $m_4$)
    \IF{$[\![\Sigma_2]\!]^\mu$ (where $\Sigma$ is the signature and it refers to a set of function symbols that define the operations or constructors available to build messages) still contains a transition with \Snd($A_3$, $l_3$, $H$, $m_3$) in its conclusion}
        \STATE skipping $m_2$ has no influence on $m_3$
        \STATE AgSt($H$, $i+1$, $\mathit{kn_{i+1}}$), Pre$_{i+1}$, \Rcv($H$, $l_3$, $A_3$, $m_3$) $\longrightarrow$ AgSt($H$, $i+2$, $\mathit{kn_{i+2}}$), Post$_{i+2}$, \Snd($H$, $l_4$, $A_4$, $m_4$)
    \ELSE 
        \STATE skipping $m_2$ has some influence on $m_3$ \COMMENT{i.e., $[\![\Sigma_2]\!]^\mu$ contains a transition with \Snd($A_3$, $l_3$, $H$, $[\![m_3]\!]^\mu$) in its conclusion}
        \STATE $\mathit{[\![kn_{i+2}]\!]^\mu}$ = $\mathit{kn_{i+1}}$ $\bigcup$ $[\![m_3]\!]^\mu$ $\bigcup$ Pre$_{i+1}$
        \STATE AgSt($H$, $i+1$, $\mathit{kn_{i+1}}$), Pre$_{i+1}$, \Rcv($H$, $l_3$, $A_3$, $[\![m_3]\!]^\mu$) $\longrightarrow$ AgSt($H$, $i+2$, $\mathit{[\![kn_{i+2}]\!]^\mu}$), Post$_{i+2}$, \Snd($H$, $l_2$, $A_2$, $m_2$) \COMMENT{as $[\![m_3]\!]^\mu$ is non-blocking towards $m_2$}
    \ENDIF
\end{algorithmic}
\end{algorithm}
\end{enumerate}

In Table~\ref{Tab:Mutations}, we instantiate the generic definition in~\ref{Eq.2} to formally define these mutations. As mentioned in Section~\ref{Sec:Introduction}, the first three mutations—\emph{skip}, \emph{add}, and \emph{replace}—are pre-existing, while \emph{disorder} is newly introduced in this paper. 

\begin{table}[tbp]
\caption{Mutations considered in this work.}
\label{Tab:Mutations}
\resizebox{\columnwidth}{!}{%
\begin{tabular}{|l|l|l|}
\hline
\multicolumn{1}{|c|}{\textbf{Notation}} & \multicolumn{1}{c|}{\textbf{Description}} & \multicolumn{1}{c|}{\textbf{Remark}}
\\ \hline
\begin{tabular}[c]{@{}l@{}} $\mu_{skip}^H$ : \\ $tr \rightarrow tr'$ \end{tabular} & \begin{tabular}[c]{@{}l@{}} [$a_0$,$\ldots$, $a_i$, $\ldots$, $a_n$]$^H$ \\ $\rightarrow$ [$a_0$,$\ldots$, $a_{i-1}$, $[\![a_{i+k}]\!]^\mu$, $\ldots$, $[\![a_n]\!]^\mu$] \end{tabular} & \begin{tabular}[c]{@{}l@{}} $H$ skips the actions \\ $a_i$,$\ldots$, $a_{i+k-1}$ \end{tabular}
\\ \hline
\begin{tabular}[c]{@{}l@{}} $\mu_{add}^H$ : \\ $tr \rightarrow tr$ \\ $\times tr'$ \end{tabular} & \begin{tabular}[c]{@{}l@{}} [$a_0$,$\ldots$, $a_i$, $\ldots$, $a_n$]$^H$ \\ $\rightarrow$ [$a_0$,$\ldots$, $a_i$, $\ldots$, $a_n$]$^H$ \\ $\times$ [$a_0$,$\ldots$, $a_{i-1}$, $[\![a_i]\!]^\mu$, $[\![a_{i+1}]\!]^\mu$, $\ldots$, \\ $[\![a_n]\!]^\mu$] \end{tabular} & \begin{tabular}[c]{@{}l@{}} Original trace $tr$ runs \\ in  parallel with the \\ new, mutated trace $tr'$ \end{tabular}
\\ \hline
\begin{tabular}[c]{@{}l@{}} $\mu_{replace}^H$ : \\ $tr \rightarrow tr'$ \end{tabular} & \begin{tabular}[c]{@{}l@{}} [$a_0$,$\ldots$, $a_i$, $\ldots$, $a_n$]$^H$ \\ $\rightarrow$ [$a_0$,$\ldots$, $[\![a_i]\!]^\mu$, $[\![a_{i+1}]\!]^\mu$, $\ldots$, $[\![a_n]\!]^\mu$] \end{tabular} & \begin{tabular}[c]{@{}l@{}} $a_i$ is a send action \\ \Snd($H$, $l$, $P$, $m$) and \\ $[\![a_i]\!]^\mu$ is its mutation \\ obtained by replacing \\ the message $m$ with \\ another message\end{tabular}
\\ \hline
\begin{tabular}[c]{@{}l@{}} $\mu_{disorder}^H$ : \\ $tr \rightarrow tr'$ \end{tabular} & \begin{tabular}[c]{@{}l@{}} [$a_0$,$\ldots$, $a_i$, $a_{i+k}$, $\ldots$, $a_n$]$^H$ \\ $\rightarrow$ [$a_0$,$\ldots$, $a_{i-1}$, $[\![a_{i+k}]\!]^\mu$, $[\![a_{i}]\!]^\mu$, $\ldots$, \\$[\![a_n]\!]^\mu$] \end{tabular} & \begin{tabular}[c]{@{}l@{}} $H$ swaps a send action \\ $a_i$ with $a_{i+k}$ \end{tabular}
\\ \hline
\end{tabular}%
}
\end{table}

Some of the example cases of \emph{skip} mutation that we identified in the UI ceremony depending on which send and/or receive actions are skipped are as follows:
\begin{itemize}[label=$-$]
    \item \emph{Case 1:} A guest $G$ does not scan the booking QR code, which corresponds to $G$ skipping \Snd(\G, $\secure$, \RK, \bookingqrcode) action.
    \item \emph{Case 2:} (As a result of Case 1) $G$ does not receive the verification link from $RK$, i.e., $G$ skipping \Rcv(\G, $\secure$, \RK, \vlink) action.
    \item \emph{Case 3:} A guest $G$ overlooks the email with the verification link, i.e., $G$ skipping \Snd(\G, $\secure$, \RK, $\langle$$\mathit{location}(\mathit{bookingqrcode})$, $\mathit{time}(\mathit{bookingqrcode})$, \vlink$\rangle$) action.
    \item \emph{Case 4:} (As a result of Case 3) $G$ does not obtain the QR code for building access, i.e., $G$ skipping \Rcv(\G, $\secure$, \RK, $\langle$\accessqrcode, \finish$\rangle$) action.
\end{itemize}

The above cases can be combined to skip bigger ``chunks'' of the ceremony execution and will be detailed in the analysis part in Section~\ref{Sec:Analysis}.
    
An example case of \emph{add} mutation is when the guest $G$ shows two QR codes (including \reservationqrcode) to the receptionist kiosk $RK$, which corresponds to $G$ executing an action \Snd(\G, $\secure$, \RK, \reservationqrcode), in addition to \Snd(\G, $\secure$, \RK, \bookingqrcode). Likewise, a case of \emph{replace} mutation is when the guest $G$ shows an incorrect QR code (\reservationqrcode), which does not match the one required by the system. This corresponds to $G$ replacing the action \Snd(\G, $\secure$, \RK, \bookingqrcode) with an action \Snd(\G, $\secure$, \RK, \reservationqrcode). Finally, a case of \emph{disorder} mutation is when the guest $G$, after obtaining the verification link (\vlink), tries to re-scan the QR code (\bookingqrcode) instead of clicking the link as expected by the ceremony.

\section{Automated Analysis with the X-Men Tool}
\label{Sec:Analysis}

In this section, we discuss how one can use the formalization in Section~\ref{Sec:Formalization} to identify attacks stemming from human agents' behavior in a security ceremony. To this end, we applied the X-Men tool to the UI ceremony. The tool generated a set of mutated models for this ceremony, which we then individually analyzed using Tamarin concerning the security goals.

\subsection{A Brief Introduction to X-Men} 
\label{Subsec:ToolDescription}

\begin{figure}[htbp]
\begin{center}
\includegraphics[width=6.5cm,keepaspectratio]{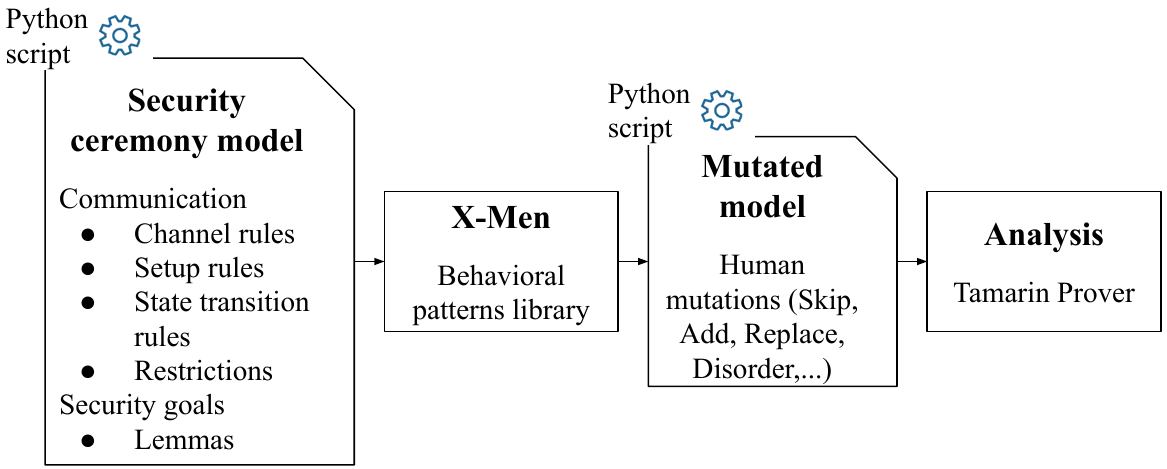}
\end{center}
\caption{Tool support architecture and approach artefacts.} 
\label{Fig:Workflow}
\end{figure}

X-Men~\cite{xmen} is a Java-based tool that extends the Tamarin Prover for modeling and analyzing security ceremonies with mutations. The tool accepts a security ceremony model as input, provided as a specification file in the \emph{.spthy} format (\emph{security protocol theory}). Then, it runs a Python script that divides the model into channel/setup/agent rules and security goals. The security analyst using X-Men can select a desired mutation and their combinations from the library of behavioral patterns to mutate the agent and other rules. The mutated parts are merged with the original channel rules and goals to produce different mutated models. These models are then used as input for analysis in Tamarin. Nevertheless, the underlying approach (cf. Fig.~\ref{Fig:Workflow}) is generic and platform-agnostic and could be applied in the context of other formal tools like ProVerif~\cite{blanchet2012automatic} and AVANTSSAR~\cite{armando2012avantssar}.

\subsection{Formal Model of the UI Ceremony} 

\emph{Channel rules} model the ceremony's initial steps, with a secure channel involved in communication between the ceremony participants. To process the ceremony roles in X-Men, they are specified inside the actions of a \emph{setup} rule. \emph{Agent rules} represent state transitions in a ceremony. A rewrite rule in Tamarin has a name and three parts, each of which is a sequence of facts: one for the rule’s left-hand side (premise), one labeling the transition (action facts), and one for the rule’s right-hand side (conclusion). Its syntax is
\begin{center}
    \textbf{rule} \emph{Rule\_name}: [\emph{Premise}] -- [\emph{Action Fact}] -$>$ [\emph{Conclusion}]
\end{center}
\emph{Restrictions} specify constraints on ceremony execution---restricting traces, behaviors, and states---on the invocation of the rules. 
\begin{paper}
Due to space constraints, we provide the Tamarin model of the UI ceremony comprising these elements in~\cite{QuamaraVigano25}, while in this section, we focus mainly on the lemmas.
\end{paper}
\begin{extended}
Due to space constraints, we provide the Tamarin model of the UI ceremony comprising these elements in Appendix~\ref{App:B} to~\ref{App:E}, while in this section, we focus mainly on the lemmas.
\end{extended}

\emph{Lemmas} specify the security properties to be proven. These are defined over traces of action facts, with each label providing sufficient information. In Tamarin, properties are specified using the following constructs:
\begin{itemize}
    \item Variables that define, for example, the roles of the ceremony, the messages exchanged between the roles, and temporal ordering between the events.
    \item Quantifiers---universal (\texttt{All}) and existential (\texttt{Ex})---define the extent to which a property holds for the variables, with temporal variables prefixed with \#;
    \item Logical operators; implication (\texttt{==$>$}), conjunction (\texttt{\&}), disjunction (\texttt{$|$}), and negation (\texttt{not});
    \item Timestamped action facts (\texttt{f @ i}, where the sort prefix for the temporal variable \texttt{`i'} is optional);
    \item (In)equalities: 
    \begin{itemize}
        \item Timestamp inequality (\texttt{i $<$ j}) for temporal ordering, where the sort prefix for the temporal variables \texttt{`i'} and \texttt{`j'} is optional,
        \item Timestamp equality (\texttt{\#i = \#j}), and
        \item Term equality (\texttt{x = y}) for message variables \texttt{`x'} and \texttt{`y'}.
    \end{itemize}
\end{itemize}

The security goals in the UI ceremony (cf. Section~\ref{Subsec:Security Goals}) are translated to Tamarin lemmas as follows:
\begin{itemize}
    \item The lemma to formalize [SG\_1] in Tamarin uses action facts to express that if the guest \texttt{G} completes the ceremony by obtaining an access QR code (\texttt{Gfin()}) at time \texttt{j}, then there is a previous instant \texttt{i} where the receptionist kiosk \texttt{RK} commits the verification link \texttt{vl} to \texttt{G} (action \texttt{CommitVerificationLink()}) (cf. Listing~\ref{Lst:Complete_VerificationLemma}).

    \begin{lstlisting}[language = xmen, caption = \emph{Complete\_Verification} lemma., label = Lst:Complete_VerificationLemma]
    lemma Complete_Verification: all-traces 
    ``(All G accessqrcode #j. Gfin(G, `qrcode', accessqrcode) @ j 
    ==> (Ex RK vl #i. CommitVerificationLink(RK, G, vl) @ i & i < j))''
    \end{lstlisting}

    \item The lemma to formalize [SG\_2] uses action facts to express that if the \texttt{G} completes the ceremony by obtaining an access QR code (\texttt{Gfin()}) at an instant in time \texttt{k} corresponding to a booking QR code (\texttt{bookingqrcode}) scanned at some previous instant \texttt{j}, then \texttt{G} did not scan any other QR code (\texttt{reservationqrcode}) at some previous instant \texttt{i} (cf. Listing~\ref{Lst:Valid_CodeLemma}).
    
    \begin{lstlisting}[language = xmen, caption = \emph{Valid\_Code} lemma., label = Lst:Valid_CodeLemma]
    lemma Valid_Code: all-traces
    ``(All G accessqrcode #k. Gfin(G, `qrcode', accessqrcode) @ k 
    ==> (Ex bookingqrcode #j. Send(G, `qrcode', bookingqrcode) @ j & j < k) 
    & not (Ex reservationqrcode #i. Send(G, `qrcode', reservationqrcode) @ i & i < k) 
    & not (reservationqrcode = bookingqrcode))''
    \end{lstlisting}

    \item The lemma to formalize [SG\_3] uses action facts to express that if \texttt{RK} commits the end of the verification (\texttt{Commit()}) at an instant \texttt{j}, and receives booking QR code (\texttt{bookingqrcode}) from the \texttt{G} and commits a verification link (\texttt{CommitVerificationLink()}) to \texttt{G} at a previous instant \texttt{i}, then there does not exist another booking QR code (\texttt{reservationqrcode}) such that \texttt{RK} execute the same transactions receiving it (cf. Listing~\ref{Lst:Transaction_ClashLemma}).
    
    \begin{lstlisting}[language = xmen, caption = \emph{Transaction\_Clash} lemma., label = Lst:Transaction_ClashLemma]
    lemma Transaction_Clash: all-traces
    ``(All G RK bookingqrcode vlink #j #i. Commit(RK, G, `finish') @ j 
    & Receive(RK, G, bookingqrcode) @ i 
    & CommitVerificationLink(RK, G, vlink) @ i &  i < j 
    ==> not (Ex reservationqrcode #l #k. 
    & Commit(RK, G, `finish') @ l 
    & Receive(RK, G, reservationqrcode) @ k 
    & CommitVerificationLink(RK, G, vlink) @ k & k < l) 
    & not (bookingqrcode = reservationqrcode))''
    \end{lstlisting}
\end{itemize}

\subsection{Analysis of the UI Ceremony} 
\label{Subsec:Analysis}

\begin{figure*}[htbp]
\begin{center}
\includegraphics[width=11cm,keepaspectratio]{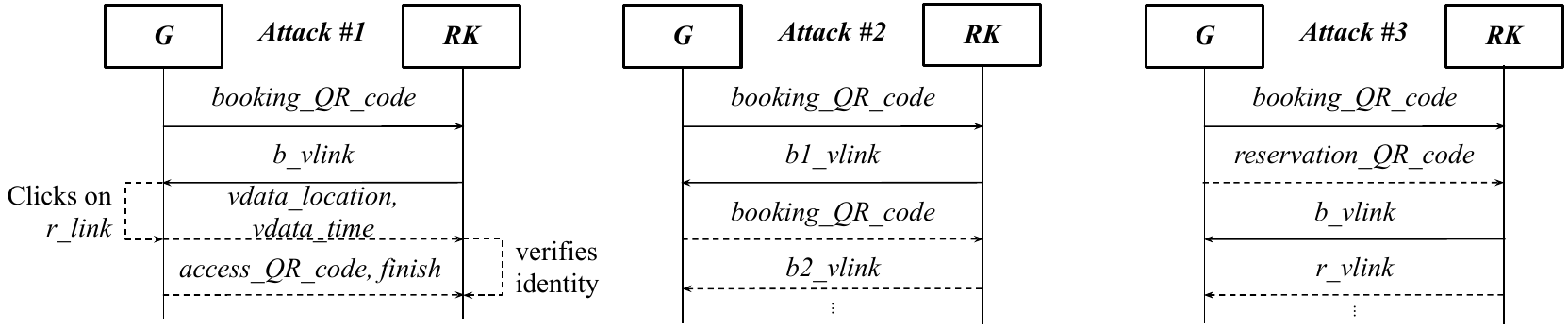}
\end{center}
\caption{Three attack scenarios for the UI ceremony.} 
\label{Fig:Attack}
\end{figure*}

Table~\ref{Tab:AttacksSummary} summarizes the attacks discovered in the UI ceremony's mutated models. Some of these attacks are described as follows (cf. Fig.~\ref{Fig:Attack}):
\begin{itemize}
    \item \emph{Attack \#1}: A guest scans a booking QR code and receives a verification link. However, they mistakenly click on a link corresponding to a different reservation, perhaps from a prior or future booking. RK sends the access QR code for this transaction and when the guest attempts to use this code for the intended meeting, the system rejects (\emph{replace} mutation causing failure of [SG\_1]). Or, if the booking system does not verify its association with the current booking, it may wrongly authenticate the guest (\emph{replace} mutation causing failure of [SG\_2]).
    \item \emph{Attack \#2}: A guest scans a booking QR code and receives a verification link. Instead of clicking the link, the guest mistakenly re-scans the booking QR code. Due to the ceremony's non-blocking nature, the system processes the re-scan as a valid action and sends another verification link to authenticate the guest, despite the out-of-order actions (\emph{disorder} mutation causing failure of [SG\_1]).
    \item \emph{Attack \#3}: If a guest inadvertently uses two booking QR codes resulting from double reservations in a repeated attempt to scan the code, it may trigger a parallel execution of the UI ceremony. This could cause RK to generate two verification links that can be used for authenticating the guest twice (\emph{add} mutation causing failure of [SG\_3]).
\end{itemize}

Another interesting case is where a guest shows a QR code corresponding to a different booking. Here, Tamarin does not detect an attack because the QR code is replaced for the entire ceremony that proceeds without indicating a problem. This exposes a vulnerability where the system could mistakenly authenticate them. To mitigate this, the system could implement a contextual check, comparing the current time with the scheduled check-in time to validate the access request. In \emph{skip} mutation cases, the ceremony remains incomplete, and the security goals cannot be verified as the ceremony is assumed to be non-blocking and the goals rely on conditional predicates. Skipping actions prevents the triggering of conditions for evaluating the implications. Thus, the system's failure to authenticate the user due to an incomplete ceremony does not violate the security goals but reflects an unmet prerequisite for their evaluation. 

\begin{table*}[tbp]
\caption{Attacks summary in mutated models [$\checkmark$: \emph{Attack found}, $\times$: \emph{Attack not found}].}
\label{Tab:AttacksSummary}
\resizebox{\textwidth}{!}{%
\begin{tabular}{|c|c|c|l|l|c|c|c|}
\hline
\textbf{Mutation} & \textbf{\multirow{2}{*}{\begin{tabular}[c]{@{}c@{}}Mutation\\Sub-type\end{tabular}}} & \textbf{\multirow{2}{*}{\begin{tabular}[c]{@{}c@{}}Mutated\\Model\end{tabular}}} &  \multicolumn{1}{c|}{\textbf{Mutation Description}} & \multicolumn{1}{c|}{\textbf{Rationale}} & \multicolumn{3}{c|}{\textbf{Security Goal}}
\\ \cline{6-8}  
& & & & & \multicolumn{1}{c|}{\textbf{SG\_1}} & \multicolumn{1}{c|}{\textbf{SG\_2}} & \multicolumn{1}{c|}{\textbf{SG\_3}}
\\ \hline
\emph{Skip} & S & M0 & Skip send in G\_1 & \begin{tabular}[c]{@{}l@{}}G does not show booking QR code to RK \end{tabular} & $\checkmark$ & $\checkmark$ & $\times$
\\ \cline{3-8}
& & M1 & Skip send in G\_3 & G does not click on the verification link & $\checkmark$ & $\times$ & $\times$
\\ \cline{2-8}
& SR & M0 & \begin{tabular}[c]{@{}l@{}} Skip send in G\_1 \\\& receive in G\_2 \end{tabular} & \begin{tabular}[c]{@{}l@{}}G does not show booking QR code to RK \\and consequently, does not receive the link \end{tabular} & $\checkmark$ & $\times$ & $\times$
\\ \cline{3-8}
& & M1 & \begin{tabular}[c]{@{}l@{}} Skip send in G\_3 \\\& receive in G\_4 \end{tabular} & \begin{tabular}[c]{@{}l@{}}G overlooks the email with the verification \\link and consequently, G does not receive \\the access QR code from RK \end{tabular} & $\times$ & $\times$ & $\times$
\\ \cline{2-8}
& R & M0 & Skip receive in G\_2 & \begin{tabular}[c]{@{}l@{}}G does not receive verification link from RK\end{tabular} & $\checkmark$ & $\times$ & $\times$
\\ \cline{3-8}
& & M1 & Skip receive in G\_4 & \begin{tabular}[c]{@{}l@{}}G does not receive access QR code from RK \end{tabular} & $\times$ & $\times$ & $\times$
\\ \cline{2-8}
& RS & M0 & \begin{tabular}[c]{@{}l@{}} Skip receive in G\_2 \\\& send in G\_3 \end{tabular} & \begin{tabular}[c]{@{}l@{}}G does not receive link from RK, consequently, \\could not share the verification data \end{tabular} & $\times$ & $\times$ & $\times$
\\ \cline{2-8}
& RSR & M0 & \begin{tabular}[c]{@{}l@{}}Skip receive in G\_2, \\ send in G\_3, and \\ receive in G\_4 \end{tabular} & \begin{tabular}[c]{@{}l@{}}G does not receive verification link from RK \\and consequently, could not share the \\ verification data to obtain access QR code\end{tabular} & $\checkmark$ & $\times$ & $\times$
\\ \hline
\emph{Add} & - & M0 & Add to send in G\_1 & G shows two booking QR codes & $\times$ & $\times$ & $\checkmark$
\\ \hline
\emph{Replace} & - & M0 & Replace send in G\_1 & \begin{tabular}[c]{@{}l@{}}G shows another booking QR code\end{tabular} & $\checkmark$ & $\checkmark$ & $\times$
\\ \cline{2-8}
& - & M1 & Replace send in G\_3 & \begin{tabular}[c]{@{}l@{}}G clicks a different/wrong verification link\end{tabular} & $\checkmark$ & $\checkmark$ & $\times$
\\ \hline
\emph{Disorder} & - & M0 & \begin{tabular}[c]{@{}l@{}}Swap send in G\_3 with \\a duplicate send in G\_1\end{tabular} & \begin{tabular}[c]{@{}l@{}}G does not click on the verification link and \\tries to re-scan the booking QR code\end{tabular} & $\checkmark$ & $\times$ & $\times$
\\ \hline
\end{tabular}%
}
\end{table*}

\section{Related Work and Positioning}
\label{Sec:RelatedWork}

Numerous works have focused on security ceremonies involving formal modeling and analysis of human errors for existing security protocol formalizations. Basin et al. in~\cite{basin2016modeling}, take a rule-based approach to formalize errors from various human behaviors, such as revealing secret information due to social-engineering attacks or entering passwords on rogue platforms, accounting for untrained, infallible, or fallible humans. They used the Tamarin Prover to analyze these formalizations against properties like device/message authentication and message secrecy in `MP-Auth' authentication protocol and phone-based protocols like `OTP over SMS'. Jacomme and Kremerin in~\cite{jacomme2021extensive}, defined a threat model for protocols involving multi-factor authentication over Transport Layer Security channels. They used Pi calculus to formalize threat scenarios, including human errors like the omission of actions, and used ProVerif to analyze the `Google 2-Step' and `U2F' protocols' versions for threat combinations. Unlike these approaches, which consider a Dolev-Yao adversary exploiting human weaknesses or protocol deviations, our analysis remains effective without such an adversary. Instead, we focus on how human errors can affect other participants' behavior and propagate throughout the ceremony. While Basin et al.~\cite{basin2016modeling} defined rules to model human errors like sending/receiving unintended messages and improper comparisons, we consider a different, yet richer, set of mutations in this work.

Statistical analysis techniques have been used to detect application-level attacks, like brute force attacks, in Virtual Reality Learning Environments~\cite{valluripally2022detection}, by considering potential user errors when typing a password for access. However, these techniques require collecting application-specific data to analyze error patterns, which are inherently probabilistic and may not offer definitive guarantees. In contrast, our work does not rely on empirical data collection. Instead, we use mathematical models and proofs to formally analyze the ceremony regarding security goals in all modeled scenarios. 

Some works have used the Tamarin Prover to analyze threats arising from human interactions in distributed settings. Bella et al. in~\cite{bella2022modelling} take an epistemic modal logic approach to formalize both intentional and erroneous human-level threats in a security ceremony. They analyzed security properties against combinations of threat models forming a lattice, using examples of Deposit-Return Systems. Their approach assumes that all technical agents behave according to the ceremony specification. In contrast, we consider the impact of human mistakes on other ceremony participants by automatically generating matching mutations with propagation rules in X-Men, producing executable traces that can be analyzed for security attacks. We adapted Sempreboni and Viganò's~\cite{sempreboni2023mutation} approach to model user identification procedures as security ceremonies, capturing human interactions with systems and agents to identify vulnerabilities arising from human errors, even without an active adversary. We also extended this approach by introducing a new mutation rule, \emph{disorder}, which permits human users to perform actions out of sequence, deviating from the original ceremony's prescribed order.

More recently, Fila and Radomirovic presented in~\cite{fila2024nothing} a formal modeling and analysis framework for ceremonies with synchronous agent interactions. The framework is generic in that it considers agents of arbitrary types, including humans, albeit in a finite number, and unifies existing symbolic models for specifying cryptographic protocols and security ceremonies. They considered the impact of execution contexts on human perception and behavior, defining them as ceremonies running parallel to the ceremony of interest. While they differentiate between normal and distracted human behavior, e.g., losing objects or forgetting information, by treating them as separate ceremonies, they offer a preliminary formalization without in-depth analysis.

\section{Conclusions}
\label{Sec:Conclusions}

We used a formal, automated approach to model and analyze user identification procedures as security ceremonies, building on and extending the work of~\cite{sempreboni2023mutation}, and applied it to a two-factor identification procedure in a real-life case study. This involved formally specifying the ceremony agents (both human and non-human), the security goals, and modeling agent mistakes as mutations, including the introduction of a new mutation, \emph{disorder}. Using this specification, we encoded the ceremony in the X-Men tool, which generated a set of mutated ceremony models that we analyzed in Tamarin Prover to identify security goal violations. Our analysis revealed vulnerabilities arising from human errors, which can affect other agents and disrupt the identification process.

We aim to extend our mutation model with timing-related mutations, where action execution delays, such as a user scanning an expired QR code or using an outdated verification link, can impact the ceremony. We also aim to perform compositional security analysis by integrating human mutations with security concerns related to other agents, such as communication processes and devices, to better comprehend complex interactions. Indeed, this would require developing strategies to tackle the increased analysis complexity. 

\section*{Acknowledgment}

This work was supported by the Horizon Europe program under the Grant Agreement 101070351 (“SERMAS: Socially-acceptable eXtended Reality Models and Systems”) and by Innovate UK.

\bibliographystyle{IEEEtran}

\newpage

\begin{extended}
\appendices

\section{Formal Model of the UI Ceremony}

\subsection{Agent Rules}
\label{App:A}

The agent rules for the receptionist kiosk role in the UI ceremony are shown in Fig.~\ref{Fig:AgentRulesReceptionistKiosk}.

\begin{figure}[htbp]
    \centering
    \resizebox{\columnwidth}{!}{
    \begin{tikzpicture}
        \node[text width=11cm, inner sep=0pt] 
        {
            \begin{itemize}[label=$RK\_\arabic$]
                \item [(RK0)] \text{[ ]}
                \\ $\xlongrightarrow{\text{$\mathit{Start}$($\mathit{RK}$, $\langle$$\langle$`$G$', `$\mathit{verificationlink}$'$\rangle$, $\langle$$G$, $\mathit{vlink}$$\rangle$$\rangle$)}}$
                \\ \text{[}AgSt($\textit{RK}$, 1, $\langle$$G$, $verificationlink$$\rangle$)\text{]}

                \item [(RK1)] \text{[}AgSt($\textit{RK}$, 1, $\langle$$G$, $verificationlink$$\rangle$), In$_{sec}$($G$, $\textit{RK}$, $\langle$`$qrcode$', $bookingqrcode$$\rangle$)\text{]} 
                \\ $\xlongrightarrow{\text{$Rcv$($\textit{RK}$, sec, $G$, $\langle$`$qrcode$', $bookingqrcode$$\rangle$)}}$
                \\ \text{[}AgSt($\textit{RK}$, 2, $\langle$$G$, $verificationlink$, $qrcode$$\rangle$)\text{]} 

                \item [(RK2)] \text{[}AgSt($\textit{RK}$, 2, $\langle$$G$, $verificationlink$, $qrcode$$\rangle$)\text{]}
                \\ $\xlongrightarrow{\substack{\text{$Snd$($\textit{RK}$, sec, $G$, $\langle$`$verificationlink$', $vlink$$\rangle$),} \\ \text{$CommitVerificationLink$($\textit{RK}$, $G$, $verificationlink$)}}}$
                \\ \text{[}AgSt($\textit{RK}$, 3, $\langle$$G$, $verificationlink$, $qrcode$$\rangle$), Out$_{sec}$($\textit{RK}$, $G$, $\langle$`$verificationlink$', $vlink$$\rangle$)\text{]} 

                \item [(RK3)] \text{[}AgSt($\textit{RK}$, 3, $\langle$$G$, $verificationlink$, $qrcode$$\rangle$), $In$($G$, $\textit{RK}$, $\langle$$\langle$`$vdata\_location$', `$vdata\_time$', `$verificationlink$'$\rangle$, $\langle$\sloppy$location(bookingqrcode)$, $time(bookingqrcode)$, $vlink$$\rangle$$\rangle$)\text{]}
                \\ $\xlongrightarrow{\substack{\text{$Rcv$($\textit{RK}$, sec, $G$, $\langle$$\langle$`$vdata\_location$', `$vdata\_time$', `$verificationlink$'$\rangle$,} \\ \text{$\langle$$location(bookingqrcode)$, $time(bookingqrcode)$, $vlink$$\rangle$$\rangle$)}}}$
                \\ \text{[}AgSt($\textit{RK}$, 4, $\langle$$G$, $verificationlink$, $qrcode$, $vdata\_location$, $vdata\_time$$\rangle$)\text{]} 
    
                \item [(RK4)] \text{[}AgSt($\textit{RK}$, 4, $\langle$$G$, $verificationlink$, $qrcode$, $vdata\_location$, $vdata\_time$$\rangle$)\text{]}
                \\ $\xlongrightarrow{\substack{\text{$Snd$($\textit{RK}$, sec, $G$, $\langle$$\langle$`$qrcode$', `$finish$'$\rangle$, $\langle$$accessqrcode$, $finish$$\rangle$$\rangle$),} \\\text{$Commit$($\textit{RK}$, $G$, `$finish$')}}}$
                \\\text{[}AgSt($\textit{RK}$, 5, $\langle$$G$, $verificationlink$, $qrcode$, $vdata\_location$, $vdata\_time$$\rangle$)\text{]}, Out$_{sec}$($\textit{RK}$, $G$, $\langle$$\langle$`$qrcode$', `$finish$'$\rangle$, $\langle$$accessqrcode$, $finish$$\rangle$$\rangle$)\text{]} 
            \end{itemize}
        };
    \end{tikzpicture}}
    \caption{Agent rules for agent $RK$ in UI ceremony.}    \label{Fig:AgentRulesReceptionistKiosk}
\end{figure}

\subsection{Channel Rules}
\label{App:B}

Secure channels are both \emph{confidential} and \emph{authentic}, preventing an adversary from reading or modifying messages or their sender. However, an attacker can still interrupt the communication by storing a message to replay later. The X-Men tool can parse channel rules specified in Listing~\ref{Lst:ChannelRules}, which can have three or four parameters depending on the \emph{types}. \texttt{SndS()}/\texttt{RcvS()} represent the fact name for a secure outbound/inbound communication. The first channel rule (\texttt{ChanSndS}) models a secure send action (\texttt{SndS()}) from participant \emph{A} to \emph{B} getting transformed into a secure channel send action (\texttt{ChanSndS()}), and then, replaced by a secure action (\texttt{!Sec()}) in the final state. The second channel rule (\texttt{ChanRcvS}) is the receive counterpart of the first. Note that to model a secure channel with an additional property to prevent replay, the persistent fact \texttt{!Sec(\$A, \$B, xn, x)} can be replaced with a linear fact \texttt{Sec(\$A, \$B, xn, x)}. Consequently, it can be used only once and cannot be replayed by an adversary at a later time.

\begin{lstlisting}[language = xmen, caption = Channel rules for the UI ceremony., label = Lst:ChannelRules]
rule ChanSndS:
[ SndS($\$$A, $\$$B, xn, x) ] 
--[ ChanSndS($\$$A, $\$$B, xn, x) ]->
[ !Sec($\$$A, $\$$B, xn, x) ] 

rule ChanRcvS:
[ !Sec($\$$A, $\$$B, xn, x) ]
--[ ChanRcvS($\$$A, $\$$B, xn, x) ]->
[ RcvS($\$$A, $\$$B, xn, x) ]
\end{lstlisting}

\subsection{Setup Rule}
\label{App:C}

We can use this rule to define each role's initial state, where the tool requires that a fact \texttt{Roles()} be instantiated. The arguments of this fact should be the agent names of the roles in the ceremony, as shown for the RK scenario in Listing~\ref{Lst:SetupRule}, where \texttt{Guest} and \texttt{RK}, are respectively the agent names of the roles guest $G$ and the receptionist kiosk $\textit{RK}$.

\begin{lstlisting}[language = xmen, caption = Setup rule for the UI ceremony., label = Lst:SetupRule]
rule Setup:
[ !Email($\$$bookingqrcode), !Link($\$$RK, $\sim$vlink) ]
--[ Setup($\$$Guest), Roles($\$$Guest, $\$$RK), RK($\$$Guest, $\$$RK) ]->
[ State($\$$RK, `1', <$\sim$vlink>), State($\$$Guest, `1', <$\$$bookingqrcode, location($\$$bookingqrcode), time($\$$bookingqrcode)>) ]
\end{lstlisting}

Additionally, if \emph{types} are needed, a \texttt{Guestsetup} rule named must be defined, with persistent facts of the form \texttt{!Type()} in its post-condition (cf.~Listing~\ref{Lst:GuestSetupRule}). Syntactically, \texttt{!Type()} includes
\begin{itemize}
    \item the agent name as the first parameter,
    \item the type as the second parameter, and
    \item the variable of that type as the third parameter.
\end{itemize}

\begin{lstlisting}[language = xmen, caption = Guest setup rule for the UI ceremony., label = Lst:GuestSetupRule]
rule Guestsetup:
[ Fr($\sim$vlink) ]
--[ OnlyOnce() ]->
[ !Type($\$$Guest, `qrcode', $\$$bookingqrcode), !Type($\$$Guest, `vdata_location', location($\$$bookingqrcode)), !Type($\$$Guest, `vdata_time', time($\$$bookingqrcode)), !Type($\$$RK, `verificationlink', $\sim$vlink), !Email($\$$bookingqrcode), !Link($\$$RK, $\sim$vlink) ]
\end{lstlisting}

\subsection{Agent Rules}
\label{App:D}

\begin{itemize}
    \item The guest in the initial state \texttt{`1'} has a booking QR code containing the meeting's location and time (cf. Listing~\ref{Lst:Guest_1Rule}). The guest then touches the mobile device at the kiosk screen, sending the code to the receptionist kiosk [\textbf{\emph{NB:}} The rules executed by a human agent (i.e., a guest) are identified using the fact \texttt{H()} as the first fact defined in the action part of each rule of the guest].

    \begin{lstlisting}[language = xmen, caption = Scanning the QR code., label = Lst:Guest_1Rule]
    rule Guest_1:
    [ State($\$$Guest, `1', <$\$$bookingqrcode, location($\$$bookingqrcode), time($\$$bookingqrcode)>) ]
    --[ H(), Send($\$$Guest, `qrcode', $\$$bookingqrcode), To($\$$RK) ]->
    [ State($\$$Guest, `2', <$\$$bookingqrcode, location($\$$bookingqrcode), time($\$$bookingqrcode)>), SndS($\$$Guest, $\$$RK, `qrcode', $\$$bookingqrcode) ]
    \end{lstlisting}
    
    \item The receptionist kiosk in the initial state \texttt{`1'} has a fresh verification link (\texttt{$\sim$vlink}) (cf. Listing~\ref{Lst:RK_1Rule}). Upon scanning the booking QR code, the receptionist kiosk sends the link to the guest’s email address associated with the booking. 

    \begin{lstlisting}[language = xmen, caption = Sending the verification link., label = Lst:RK_1Rule]
    rule RK_1:
    [ State($\$$RK, `1', <$\sim$vlink>), RcvS($\$$Guest, $\$$RK, `qrcode', $\$$bookingqrcode) ]
    --[ Receive($\$$RK, $\$$Guest, $\$$bookingqrcode), CommitVerificationLink($\$$RK, $\$$Guest, $\sim$vlink) ]-> 
    [ State($\$$RK, `2', <$\sim$vlink, $\$$Guest, $\$$bookingqrcode>), SndS($\$$RK, $\$$Guest, `verificationlink', $\sim$vlink) ]
    \end{lstlisting}
    
    \item Upon receiving the link, the guest confirms their identity by clicking it, thereby sending the verification data to the receptionist kiosk (cf. Listing~\ref{Lst:Guest_2Rule}).
    
    \begin{lstlisting}[language = xmen, caption = Sending the verification data., label = Lst:Guest_2Rule]
    rule Guest_2:
    [ State($\$$Guest, `2', <$\$$bookingqrcode, location($\$$bookingqrcode), time($\$$bookingqrcode)>), RcvS($\$$RK, $\$$Guest, `verificationlink', $\sim$vlink) ]
    --[ H(), Receive($\$$Guest, $\$$RK, $\sim$vlink), Send($\$$Guest, `vdata_location', location($\$$bookingqrcode)), Send($\$$Guest, `vdata_time', time($\$$bookingqrcode)), Send($\$$Guest, `verificationlink', $\sim$vlink), To($\$$RK) ]->
    [ State($\$$Guest, `3', <$\$$bookingqrcode, location($\$$bookingqrcode), time($\$$bookingqrcode), $\sim$vlink>), SndS($\$$Guest, $\$$RK, <`vdata_location', `vdata_time', `verificationlink'>, <location($\$$bookingqrcode), time($\$$bookingqrcode), $\sim$vlink>) ]
    \end{lstlisting}
    
    \item The receptionist kiosk 
    sends the building access QR code, along with a \texttt{finish} flag to the guest (cf. Listing~\ref{Lst:RK_2Rule}).
    \begin{lstlisting}[language = xmen, caption = Sending the access QR code., label = Lst:RK_2Rule]
    rule RK_2:
    [ State($\$$RK, `2', <$\sim$vlink, $\$$Guest, $\$$bookingqrcode>), RcvS($\$$Guest, $\$$RK, <`vdata_location', `vdata_time', `verificationlink'>, <location($\$$bookingqrcode), time($\$$bookingqrcode), $\sim$vlink>) ]
    --[ Receive($\$$RK, $\$$Guest, location($\$$bookingqrcode)), Receive($\$$RK, $\$$Guest, time($\$$bookingqrcode)), Commit($\$$RK, $\$$Guest, `finish') ]->
    [ State($\$$RK, `3', <$\sim$vlink, $\$$bookingqrcode, location($\$$bookingqrcode), time($\$$bookingqrcode)>), SndS($\$$RK, $\$$Guest, <`qrcode', `finish'>, <accessqrcode, `finish'>) ]
    \end{lstlisting}
    
    \item The guest receives the access QR code on their email to access the building, marking the end of the UI ceremony (cf. Listing~\ref{Lst:Guest_3Rule}). 
    \begin{lstlisting}[language = xmen, caption = Receiving the access QR code., label = Lst:Guest_3Rule]
    rule Guest_3:
    [ State($\$$Guest, `3', <$\$$bookingqrcode, location($\$$bookingqrcode), time($\$$bookingqrcode), $\sim$vlink>), RcvS($\$$RK, $\$$Guest, <`qrcode', `finish'>, <accessqrcode, `finish'>) ]
    --[ H(), Gfin($\$$Guest, `qrcode', $\$$accessqrcode) ]->
    []
    \end{lstlisting}
\end{itemize}

\subsection{Restrictions}
\label{App:E}

\begin{itemize}
    \item The restriction in Listing~\ref{Lst:UniqueRole} enforces a structure where the participant \texttt{G1} has a distinct role than \texttt{RK1} and \texttt{RK2} in the ceremony. However, participants \texttt{G1} and \texttt{G2} are essentially playing the same role, i.e., the guest, in the ceremony.
    \begin{lstlisting}[language = xmen, caption = \emph{UniqueRole} restriction., label = Lst:UniqueRole]
    restriction UniqueRole:
    ``All G1 G2 RK1 RK2 #i #j. Roles(G1, RK1) @ i & Roles(G2, RK2) @ j 
    ==> not G1 = RK1 & not G1 = RK2 & G1 = G2''
    \end{lstlisting}

    \item The restriction in Listing~\ref{Lst:OnlyOnceRestriction} ensures the guest can perform only one verification session in a ceremony. To enforce this, \texttt{OnlyOnce()} is used as an action fact in the rule and this restriction is added.
    \begin{lstlisting}[language = xmen, caption = \emph{OnlyOnce} restriction., label = Lst:OnlyOnceRestriction]
    restriction OnlyOnce:
    ``All #i #j. OnlyOnce() @ #i & OnlyOnce() @ #j 
    ==> #i = #j''
    \end{lstlisting}
\end{itemize}

\subsection{Lemma}
\label{App:F}

In Listing~\ref{Lst:FunctionalLemma}, the universal quantification part suggests that the setup events for \texttt{G1} and \texttt{G2} must occur at the same index. The existential quantification part indicates that if both the sub-parts occur at index \texttt{k} and \texttt{l}, respectively, there exists a trace that satisfies these conditions. 

\begin{lstlisting}[language = xmen, caption = \emph{Functional} lemma., label = Lst:FunctionalLemma]
lemma functional: exists-trace
``(All G1 G2 #i #j. Setup(G1) @ i & Setup(G2) @ j 
==> #i = #j) 
& (Ex G bookingqrcode RK #k #l. Gfin(G, `qrcode', bookingqrcode) @ k
& Commit(RK, `Guest', `finish') @ l)''
\end{lstlisting}
\end{extended}

\end{document}